# A Model of Triple-Channel Interaction Dynamics in Pharmaceutical Retailing in Emerging Economies

Koushik Mondal[a*], Balagopal G Menon[b], Sunil Sahadev[c]

[a, b] Department of Industrial and Systems Engineering,
Indian Institute of Technology Kharagpur, Kharagpur 721 302, West Bengal, India.

[c] Department of Management, Sheffield Hallam University,
S11WA Sheffield, United Kingdom.

**Abstract**

The survival of unorganized pharmacies is increasingly challenging in the face of growing competition from organized and e-pharmaceutical retail channels in emerging economies. A theoretical model is developed to capture the triple-channel interactions among unorganized, organized and e-retailing in emerging markets, taking into account the essential features of the pharmaceutical retail landscape, consumers, retailers and pharmaceutical products. Given the retailer and customer-specific factors, the price-setting game between the triple-channel retailers yielded the optimal prices for these retailers. The analysis found that the product category level demand has no influence on optimal pricing strategies of the retailers. The analysis also reveals counterintuitive results, for instance, (*i*) an increase in customer acceptance of unorganized retailers will result in a decrease in profits of both unorganized and organized retailers; (*ii*) as the distance and transportation cost to unorganized retailers increases for the consumers, the profit of the unorganized retailer increases; and (*iii*) consumers' marginal utility of money has no influence on the optimal price, but have an influence on the profit of the three retail channels. Our research findings offer valuable insights for policymakers facing challenges in achieving a balanced growth among the organized, unorganized, and e-pharmaceutical retail sectors in emerging economies.

## 1. Introduction

The retail industry serves as the final member of the supply chain that directly faces customer demand fluctuations and, by fulfilling them, generates cash flow which is the lifeline of all supply chain members (Chopra and Meindl 2007). The retail sector constitutes a significant share of most national economies. In the U.S., retail sales topped $6.5 trillion in 2021 (U.S. Census Bureau 2022), and in emerging markets like India they reached about $850 billion, roughly 30 percent of GDP of India (Technopak 2020). The retail sector is generally divided into organized and unorganized retailing. Organized retail covers the large, licensed players like supermarkets, hypermarkets, and national chains. Whereas unorganized retail refers to traditional local family-owned mom & pop shops often called as '*kirana*' in India, '*spaza*' in South Africa, '*changarro*' in Mexico, '*sari-sari*' in the Philippines, '*bodega*' in Peru, '*warung*' in Indonesia (Escamilla et al. 2021). In literature, these space and cash-constrained, family-run stores sometimes called nanostores (Blanco and Fransoo 2013), micro-retailers (Zhang et al. 2017) or small unorganized retailers (Dugar and Chamola 2021), thrive on low overhead costs, tight-knit customer ties and convenient locations. They serve a few hundred regular customers, often extend relationship-based credit, personalized service and even hand-deliver orders (Child et al. 2015). Across South Asia and Africa, they handle over 85 percent of consumer-packaged goods; in Latin America, about half the market (Fransoo et al. 2017). No surprise, then, that there are an estimated 50 million nanostores in developing nations; 12 million in India alone (Kohli and Bhagwati 2012), six million in China (Ge 2017), plus millions more from Mexico to Nigeria and other developing economies.

Foreign direct investment (FDI) liberalization policies in developing countries (for example India, Chilie, Mexico, Brazil, South Africa etc.) and the prospects of exponential growth have prompted global organized retailing giants such as Walmart, Costco, and Amazon to venture into the emerging markets (Alexander and Myers 1999, Jerath et al. 2016). Furthermore, national conglomerates and large corporations are proactively expanding their retail franchises in their native countries, fuelling the organic growth of organized retailing in developing markets (Jerath et al. 2016). Such growth in the retail industry has significant implications for the wider economy, including job creation, infrastructure development, and increased consumer spending (Alexander and Myers 1999). Despite the conventional view that organized retailing brings efficiency into the market, there have been reports of adverse effects like conflicts with local existing small unorganized retailers and deterioration of the social fabric (Bianchi and Ostale 2006, Mireles 2005, Singh et al. 2006). Nevertheless, the organized retail as well as online e-retail sector are expanding in emerging countries despite of these negative sentiments (Alexander and de Lira e Silva 2002, Maharajh and Heitmeyer 2005). Despite of the significance of this phenomenon, the impact of organized and e-retailers on unorganized retailers in emerging economies is still not fully understood (Jerath et al. 2016).

Figure 1. Organized, Unorganized and E-Pharmaceutical Pharmaceutical Supply Chain

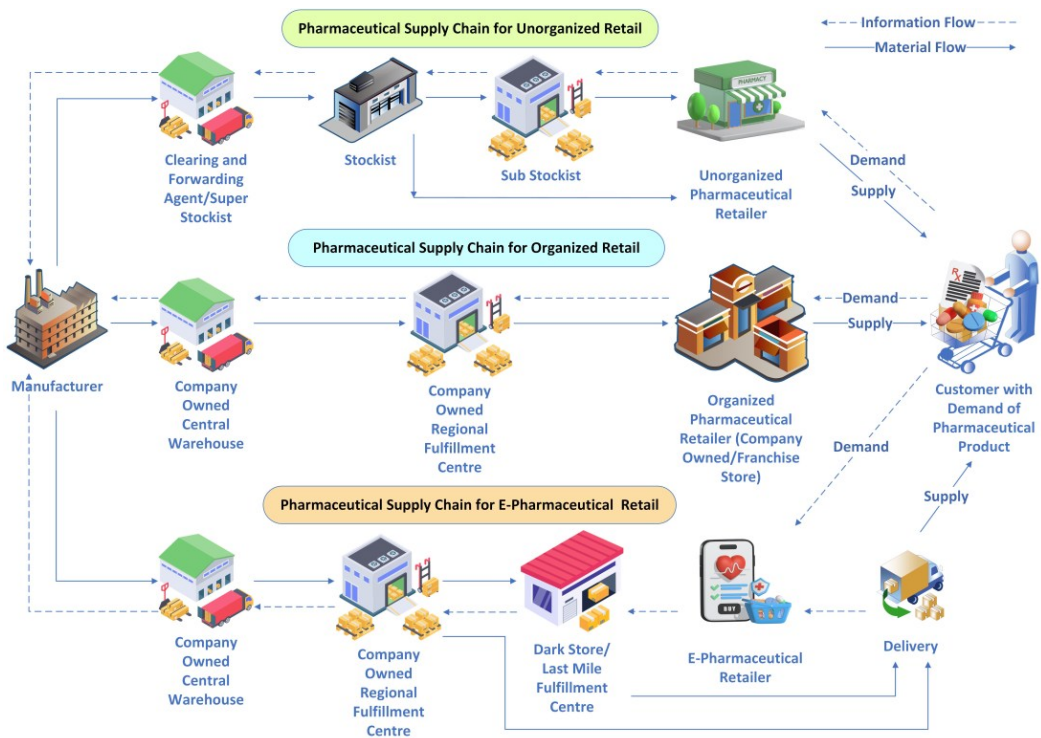

Community pharmacies are also called pharmaceutical retail, bridge physician prescriptions and patient care, staffed by licensed pharmacists who dispense, verify, and counsel on medications and wellness (Berenbrok et al. 2022). Seminal studies by Yadav (2015) and Larson et al. (2013) have exposed supply chain key issues and needed reforms in developing economies and proposed distribution and financing reforms. Today's pharmaceutical retail outlets range from space constrained, individually owned outlets in urban and rural neighbourhoods to large chain stores and online e-pharmaceutical retailers, reflecting their growing role in patient-centred medication management. Figure 1 represents a typical organized, unorganized and e-pharmaceutical supply chain in India.

Recent projections of the Indian Pharmaceutical Alliance (IPA 2024) reveal that the valuation of the Pharma industry will reach around INR 4.60 lakh crores (US $55 billion) by 2030 from INR 2.42 lakh crores (US $29 billion) in 2023, supported by an estimated compound annual growth of 9.6%. Notably, digitalization of the healthcare sector has boosted the growth of the e-pharmacy segment, which is set to increase considerably around INR 24,000 crores by 2030 from INR 3,800 crores in 2019, while the organized pharmacy chains are set to gain a larger market presence, with revenues forecasted to increase around INR 37,000 crores by 2026 from INR 13,400 crores in 2020 (Market Research 2022, IBEF 2024). Nonetheless, unorganized retail remains predominant, rising from INR 1,54,800 crores in 2020 to INR 2,39,000 crores by 2026 (IBEF 2024, Market Research 2022). Accounting for roughly 54 percent of India's pharmaceutical retail market, the unorganized segment continues to expand at about 10 percent annually (IPA 2024) yet faces mounting pressures. Organized chains have deployed aggressive tactics like minimum 20 percent discounts, loyalty programmes and broader product assortments, that the traditional outlets struggle to match. As a result, the unorganized sector's share has fallen from 62.5 percent in 2019 to 54.2 percent in 2023 (IPA 2024, IBEF 2024, Market Research 2022).

Motivated by the above, the present study aims to answer specific questions: (1) What factors influence customers' choice of purchasing between unorganized, organized, and e-pharmaceutical retail channels? (2) How does a customer's selection of a specific pharmaceutical retail channel influence the pricing strategies and profitability of unorganized, organized, and e-pharmaceutical retailers? (3) What are the conditions for achieving equilibrium in a market where organized, unorganized and e-pharmaceutical retail channels co-exist? The study focuses exclusively on unorganized, organized, and e-pharmaceutical retailers, limiting its scope to the pharmaceutical retail sector in emerging economies like India.

## 2. Literature Review

The present study relates to four key streams of literature namely price competition among spatially distributed retailers, demand modeling incorporating consumer heterogeneity, retail operations within emerging markets, and the dynamics of multi-channel competition in the pharmaceutical retail sector.

The literature on spatial competition is vast and widely explored in academic research, starting from two foundational models which are the Hotelling linear model (Hotelling 1929) and the Salop circle model (Salop 1979). Biscaia and Mota (2013) provide a comprehensive review of the same. A number of studies have considered symmetrically distributed identical organized retailers and an organized retailer at the centre of a Salop circle. For example, Balasubramanian (1998), Jerath et al. (2016), and Ge and Tomlin (2025). Consumer purchasing behavior critically influences competition among unorganized, organized and e-pharmacies. Consumers face a trade-off between frequent purchases at local unorganized outlets, bulk buying from organized retailers with higher transaction costs, and home delivery from online platforms with attendant waiting times (Jerath et al. 2016). Jerath et al.'s (2016) model indicates that entry by an organized retailer can reduce both consumer and social surplus, though it abstracts from asymmetric retailer locations, multiple organized entrants, extended time horizons and product assortment differences. Their model, however, assumes perfectly symmetric locations for small unorganized shops, includes just one organized entrant, spans only two periods and ignores variations in product assortments.

Ge and Tomlin (2025) use a multi-party supply-chain game to analyze how wholesale-price contracts shape competition between supermarkets and nanostores. They show that as nanostore density rises, both aggregate demand and profits in that channel can fall, while the supermarket's market power boosts consumer surplus. Like Jerath et al. (2016), their model relies on a symmetric "circle" layout for unorganized retailers and omits dynamic entry and exit decisions by nanostores.

In demand modeling of multi-channel retail competition, most of the studies have assumed a constant demand. Chiang et al. (2003) propose a customer valuation approach where demand is derived using the indifference point of customer valuation of purchasing from a particular retail channel. They have studied how the customer acceptance of online channels, as a substitute for traditional in-store shopping, can shape the design of the supply chain. Whereas Jearth et al. (2016) have proposed a disutility along with product valuation, Rofin and Mahanty (2018) have considered price-sensitive customers using money utility. Our work progresses this research further by introducing customer acceptance of unorganized retail channels along with price-sensitive heterogeneous customers and the disutility of purchasing from the organized and e-pharmaceutical retailer in a single framework.

Several studies have explored diverse facets of retail operations in emerging markets. Narayan et al. (2015) studied the adoption of organized retailing in India and find that consumers in the upper and lower middle classes is the most responsive to organized retailing. Zhang (2015) examined how companies choose to rebrand in ways that distance themselves from their emerging market roots in order to reduce negative perceptions. Gui et al. (2019) and Zhang et al. (2017) explore novel replenishment strategies for rural nanostores, whilst Iyer and Palsule-Desai (2019) analyze the functions of intermediaries in the stocking and distribution processes for Indian nanostores. Ge et al. (2021) examine a manufacturer's distribution strategy for these retail outlets, whereas Ge et al. (2020) study competition among manufacturers for constrained shelf space in the nano-retail channel. Escamilla et al. (2021) and Fransoo et al. (2024) provide an extensive review of nanostore-related academic research.

Pharmaceutical literature has examined the competition between large organized chain pharmaceutical retailers and small unorganized retailers (Miller & Goodman 2017), highlighted critical role of community pharmacies in rural areas (Berenbrok et al. 2022), and explored pharmacists' contributions to chronic disease management (Mossialos et al. 2015). Yet these studies consistently view the landscape through a practitioner's lens, leaving the managerial perspective underexplored.

In this present study, we build a theoretical model to capture the triple-channel interactions among unorganized, organized and e-retailing in developing countries by considering essential characteristics of the pharmaceutical retailing environment, the consumers, the retailers and the pharmaceutical products. We provide analytical results that shed light on the retailers' behaviour in response to various influential retailer and customer-specific factors. This study offers several interesting contributions to the field of operations management, particularly in relation to the pharmaceutical retail industry. First, unlike previous research in the retail sector that employed a 'circle model' assuming unorganized retailers located in the market in a perfectly symmetrical manner (Jerath et al. 2016, Ge and Tomlin 2025), our study does not adopt such assumptions of symmetry to customer and retailer locations. Recognizing geographical limitations may hamper perfectly symmetrical positioning in reality (Jerath et al. 2016), our study makes the model closer to real-world market scenarios. Second, unlike taking a single organized retailer in the market, as in Jerath et al. (2016) and Ge and Tomlin (2025), our study considers multiple unorganized, organized, and e-pharmaceutical retailers along with spatially distributed customers to model the market dynamics more realistically. Third, the existing studies (Chiang et al. 2003, Jerath et al. 2016) have considered 'identical customers', whereas in our study, diversified price-sensitive customers (denoted as '$m$') are considered. Chiang et al. (2003) introduced the factor "customer acceptance of online retail channel"; and in the present study we have introduced an additional factor '$\alpha$' which is the "customer acceptance of an unorganized pharmaceutical retailer". Therefore, this study is among the early contributions to the field to integrate three types of retailer channels into one framework, while also considering customer behavior dynamics in India's unorganized pharmaceutical retail sector.

## 3. Model

### 3.1. Overview

A game-theoretical model is developed to capture the complex market dynamics where unorganized, organized, and e-pharmaceutical retailers coexist in a developing country like India. The model is initially developed based on individual customers' channel choice behavior. From the literature and the field study, several factors are identified that influence customers' channel choice behavior, and subsequently, the utility functions of the customers are proposed. Each customer purchases from that pharmaceutical retail channel, which provides maximum utility. Demand for each retail channel is derived analytically by assuming customer valuation as uniformly distributed within the consumer population from 0 to 1. The profit functions of the retailers are proposed, and optimal price and profits are calculated by constructing a price-setting Bertrand game among the unorganized, organized, and e-pharmaceutical retailers. After that, numerical simulation and sensitivity analysis are carried out to check the robustness of the model. In this section, the characteristics of pharmaceutical retailing environments, customers, retailers, and product characteristics are discussed, along with assumptions and notations used in the model, which are presented in Table 1.

Table 1. Notations Used in the Model

| Notations Parameter | Definition | References |
|---|---|---|
| $c_1$ | Marginal cost incurred by the unorganized pharmaceutical retailers (Rs.) | Chiang et al. (2003), Jerath et al. (2016) |
| $c_2$ | Marginal cost incurred by the organized pharmaceutical retailers (Rs.) | Chiang et al. (2003), Jerath et al. (2016) |
| $c_3$ | Marginal cost incurred by the e-pharmaceutical retailers (Rs.) | Chiang et al. (2003), Jerath et al. (2016) |
| $v$ | Customer valuation of the product ($0 \leq v \leq 1$) (Rs.) | Chiang et al. (2003) |
| $\beta$ | Probability of product category level demand ($0 \leq \beta \leq 1$) | Jerath et al. (2016) |
| $t$ | Transportation cost per unit distance incurred by the customer (Rs. /km) | Jerath et al. (2016) |
| $x$ | Average distance to the nearest unorganized pharmaceutical retailer for a customer (km) | Jerath et al. (2016) |
| $\theta$ | Customer acceptance of e-pharmaceutical retailer ($0 \leq \theta \leq 1$) | Chiang et al. (2003), Rofin and Mahanty (2018) |
| $\alpha$ | Customer acceptance of an unorganized pharmaceutical retailer ($0 \leq \alpha \leq 1$) | Present study |
| $m$ | Marginal utility of money ($0 \leq m \leq 1$) | Rofin and Mahanty (2018) |
| $\mu_1$ | Customer disutility of purchasing from the organized pharmaceutical retailer (Rs.) | Jerath et al. (2016) |
| $\mu_2$ | Customer disutility of purchasing from the e-pharmaceutical retailer (Rs.) | Present study |
| **Notations Used in Equation** | | |
| $U_u$ | Utility derived from the unorganized pharmaceutical retailer (Rs.) | |
| $U_o$ | Utility derived from the organized pharmaceutical retailer (Rs.) | |
| $U_e$ | Utility derived from the e-pharmaceutical retailer (Rs.) | |
| $D_u$ | Demand for the unorganized pharmaceutical retailers | |
| $D_o$ | Demand for the organized pharmaceutical retailers | |
| $D_e$ | Demand for the e-pharmaceutical retailers | |
| $\pi_1$ | Profit of the unorganized pharmaceutical retailers (Rs.) | |
| $\pi_2$ | Profit of the organized pharmaceutical retailers (Rs.) | |
| $\pi_3$ | Profit of the e-pharmaceutical retailers (Rs.) | |
| $v^u$ | Indifferent customer valuation point when purchasing from the unorganized pharmaceutical retailers | |
| $v^o$ | Indifferent customer valuation point when purchasing from the organized pharmaceutical retailers | |
| $v^e$ | Indifferent customer valuation point when purchasing from the e-pharmaceutical retailers | |
| $v^{uo}$ | Indifferent customer valuation point when purchasing from the unorganized and organized pharmaceutical retailers | |
| $v^{oe}$ | Indifferent customer valuation point when purchasing from the organized and e-pharmaceutical retailers | |
| $v^{ue}$ | Indifferent customer valuation point when purchasing from the unorganized and e-pharmaceutical retailers | |
| $v^{uoe}$ | Indifferent customer valuation point when purchasing from the unorganized, organized, and e-retailers | |
| **Decision Variables** | | |
| $p_1$ | Price charged by the unorganized pharmaceutical retailers (Rs.) | Jerath et al. (2016), Pazgal et al. (2013) |
| $p_2$ | Price charged by the organized pharmaceutical retailers (Rs.) | Jerath et al. (2016), Pazgal et al. (2013) |
| $p_3$ | Price charged by the e-pharmaceutical retailers (Rs.) | Jerath et al. (2016), Pazgal et al. (2013) |

## 3.2. The Consumers

Customers are assumed to be rational. They want to maximize their utility by comparing the price and disutility of purchasing from a retail channel with the perceived value of the product, before making the purchasing decision. In a multi-channel market, customers can buy a pharmaceutical product from either the traditional independent unorganized pharmaceutical retailer, the organized chain pharmaceutical retailer, or the online e-pharmaceutical retailer.

Consumers are assumed to be heterogeneous in their assessment of the product. In order to simplify the analysis, we denote the consumption value (alternatively referred to as "willingness to pay") as '$v$'. We presume that it is uniformly distributed within the consumer population from 0 to 1, with a density of 1. Let customer valuation of the product be $v$ when bought through an organized retailer, $\alpha v$ when purchased from an unorganized retailer, and '$\theta v$' when purchased through an online retailer, where $0 < \alpha < 1$ and $0 < \theta < 1$. The value of the parameter '$\theta$' is called the customer acceptance of an online retail channel, and the parameter '$\alpha$' is called the customer acceptance of an unorganized pharmaceutical retail channel. It is assumed that the customer valuations of products for online e-pharmaceutical retailers and unorganized pharmaceutical retailers are less than those of the organized pharmaceutical retailers. This is because of two shortcomings of online purchase (Chiang et al. 2003, Liang & Huang 1998, Kacen et al. 2013): physical examination of the product and immediate possession of the product are not possible. On the other hand, the shortcomings of purchasing from an unorganized pharmaceutical retailer are a lesser variety of assortment due to budget, and storage constraints of the unorganized independent pharmaceutical retailers (Fransoo et al. 2017), which may lead to frequent stock-outs. Especially for emergency pharmaceutical products, customers want to fill their whole prescriptions as quickly as possible, preferably from a single store with a higher assortment to minimize the travel time during emergencies.

We have assumed a linear demand function similar to Chiang et al. (2003) and Jerath et al. (2016). However, we have modified the demand function to make it more realistic by considering additional parameters, such as money utility ($m$), customer disutility of purchasing from the organized pharmaceutical retailer ($\mu_1$), and customer disutility of purchasing from the e-pharmaceutical retailer ($\mu_2$). Further, the demand function captures the effect of customer acceptance of the online channel ($\theta$) and the customer acceptance of the unorganized retailer channel ($\alpha$), which are highly dynamic, especially in developing countries like India. The factor '$\alpha$' is not considered for organized and e-pharmacies because the assortments of organized and e-pharmacies are larger than unorganized retailers (Blanco and Fransoo 2013). The factor is important because consumers rank variety of assortment right behind location and price when naming reasons why they patronize their favourite stores (Dukes et al. 2009, Hoch et al. 1999, Krishnan et al. 2002).

### 3.3. Characteristics of the Retailers

There are three types of retail present in the market i.e., unorganized, organized and e-pharmaceutical retailers. All the retailers sell identical homogeneous pharmaceutical products. Marginal costs incurred by the three different retailers are different due to their different modes of operation. Organized, unorganized, and e-pharmaceutical retailer channels serve customers in completely different ways (Fransoo et al. 2017). The profit and market share of these three pharmaceutical retail channels depend on the customer demand for each of these three channels (Jerath et al. 2016). Figure 2 shows the difference notations for the triple-channel pharmaceutical supply chains.

Figure 2. Triple-Channel Pharmaceutical Supply Chain

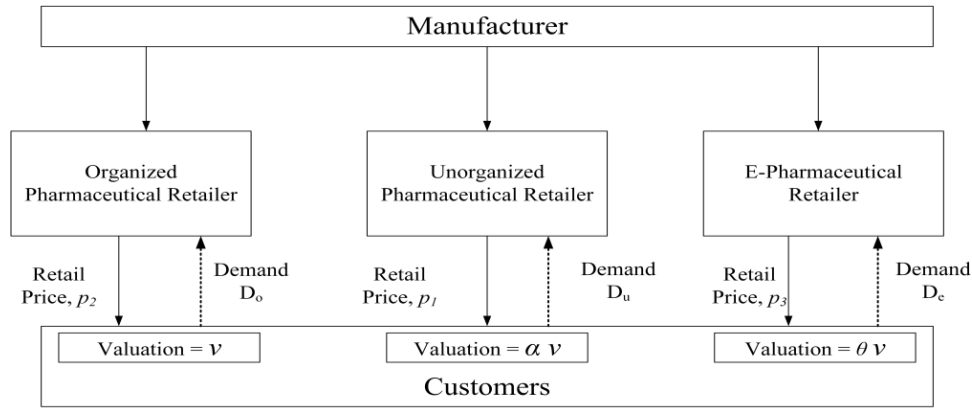

### 3.4. Characteristics of the Pharmaceutical Product

In pharmaceutical retailing, drugs are broadly classified as prescription drugs and non-prescription drugs or over-the-counter (OTC) drugs. Customers generally seek a physician and purchase the medicines from any of the three retail channels as prescribed by the doctors. Therefore, unlike in grocery shopping, here the customer is not indifferent between whether to buy each time when the demand occurs (Buy to consume) or buy more units together in anticipation of future demands (Buy to stock) (Jerath et al. 2016). That's why in this model; each customer is modelled to buy when the demand occurs and the storage capability factor is automatically waived off. The argument behind this is apart from some very emergency products (like vaccine, plasma), most of the pharmaceutical drugs require very small storage space or specific storage conditions for customers. We have also assumed a probability of product category level demand $\beta$ for each customer (Jerath et al. 2016). The parameter $\beta$ varies with product category but stays constant over time. For some pharmaceutical product categories, $\beta$ is large (e.g., drugs for chronic diseases, seasonal diseases), whereas for other categories $\beta$ may be small (e.g., emergency rare drugs).

### 3.5. Game Framework

The market competition among the three retail distribution channels is modelled using a single-period Bertrand Price competition game. Bertrand competition is a model wherein two or more firms supply a homogeneous product to the customer and participate in price competition. The pharmaceutical products are assumed homogeneous because they are perfect substitutes due to the fact that the composition of drugs of two different brands is the same. The three players, i.e., the unorganized, organized, and e-pharmaceutical retailers, do not compete in quantities but in prices. Each retailer decides the price of its product in order to maximize its profit, while considering the pricing offered by other retailers. This makes Bertrand price competition perfect for the present game theoretical model.

# 4. The Retail Pricing Game

## 4.1. Derivation of Demand Functions

Firstly, we formulate the demand functions for unorganized, organized, and e-pharmaceutical retailers and derive the respective profit functions.

The utility of a customer located at a distance $x$ and purchasing from the nearest unorganized retailers is given by: $U_u = \alpha v\text{-}mp_1\text{-}tx$. The condition for customers to purchase from unorganized pharmaceutical retailers is $U_u > 0$, which leads to $v > (mp_1+tx)/\alpha$. The scenario in which a customer is indifferent to purchase from the unorganized retailers or not is $U_u = 0$. The indifference point $v^u$ is given by: $v^u = (mp_1+tx)/\alpha$. Similarly, utility derived from purchasing from the organized pharmaceutical retailers is: $U_o = v\text{-}mp_2\text{-}\mu_1$. The condition under which a customer is indifferent to purchase from the organized pharmaceutical retailer or not is $U_o = 0$. The indifference point $v^o$ is given by: $v^o = mp_2+\mu_1$. Similarly, utility derived from purchasing from the e-pharmaceutical retailer is: $U_e = \theta v\text{-}mp_3\text{-}\mu_2$. The condition under which a customer is indifferent between purchasing from an online e-pharmaceutical retailer or not is $U_e = 0$. The indifference point $v^e$ is given by: $v^e = (mp_3+\mu_2)/\theta$. Next, different market scenarios are described to find out the parameter conditions under which all three retail channels will have demand.

### 4.1.1. Market Where Organized Retailers and E-Pharmaceutical Retailers are Present:

When customers have the option to purchase from an organized or e-pharmaceutical retailer, they will compare $U_o$ and $U_e$. If $U_o > U_e$, then the organized retail is preferred over e-pharmaceutical retail and if $U_o = U_e$, the customer is indifferent between purchasing from organized and e-pharmaceutical retailers. The indifference customer valuation points $v^{oe}$ is given by the following indifference equation:

$$v^{oe} - mp_2 - \mu_1 = \theta v^{oe} - mp_3 - \mu_2 \tag{1}$$

Which simplifies to;

$$v^{oe} = \frac{m(p_2 - p_3) + \mu_1 - \mu_2}{(1 - \theta)} \tag{2}$$

Note: if the valuation exceeds $v^{oe}$, customer will prefer the organized retailer over e-pharmaceutical retailer.

Now we have two cases: $v^e < v^o$ and $v^e > v^o$.

**Case 1:** when $\nu^e < \nu^o$,

$$\frac{mp_3 + \mu_2}{\theta} < mp_2 + \mu_1 \tag{3}$$

$$-\theta(mp_2 + \mu_1) < -(mp_3 + \mu_2) \tag{4}$$

$$(mp_2 + \mu_1) - \theta(mp_2 + \mu_1) < (mp_2 + \mu_1) - (mp_3 + \mu_2) \tag{5}$$

$$mp_2 + \mu_1 < \frac{m(p_2 - p_3) + \mu_1 - \mu_2}{(1 - \theta)} \tag{6}$$

Or, $\nu^o < \nu^{oe}$, thus $\nu^e < \nu^o < \nu^{oe}$. The situation is illustrated in Figure 3.

Figure 3. Customer Valuation When $\nu^e < \nu^o$

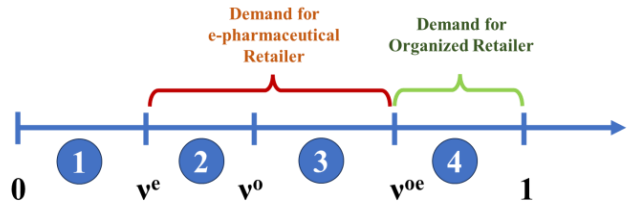

The continuous scale of customer valuation is divided into four zones, namely 1, 2, 3 and 4. For customers in zone 1, both utilities from organized and e-pharmaceutical retailer are less than zero. Therefore, they will not buy from either channel. In zone 2, utility from the e-pharmaceutical retailer $\geq 0$ and it is more than the utility from organized retailer in zone 3, so customers will buy from e-pharmaceutical retailer in these two sections. Customers will buy from the organized retailer only when their valuation exceeds the customer indifference point $\nu^{oe}$ in zone 4 where $U_o > U_e$.

Demand for organized retailer is:

$$D_o = 1 - \nu^{oe} \tag{7}$$

Demand for e-pharmaceutical retailer is:

$$D_e = \nu^{oe} - \nu^e \tag{8}$$

**Case 2:** when $\nu^e > \nu^o$,

$$\frac{mp_3 + \mu_2}{\theta} > mp_2 + \mu_1 \tag{9}$$

$$-(mp_3 + \mu_2) < -\theta(mp_2 + \mu_1) \tag{10}$$

$$(mp_2 + \mu_1) - (mp_3 + \mu_2) < (mp_2 + \mu_1) - \theta(mp_2 + \mu_1) \tag{11}$$

$$\frac{m(p_2 - p_3) + \mu_1 - \mu_2}{(1 - \theta)} < mp_2 + \mu_1 \tag{12}$$

Or, $\nu^{oe} < \nu^o$, thus $\nu^{oe} < \nu^o < \nu^e$. The situation is illustrated in Figure 4.

Figure 4. Customer Valuation When $\nu^e > \nu^o$

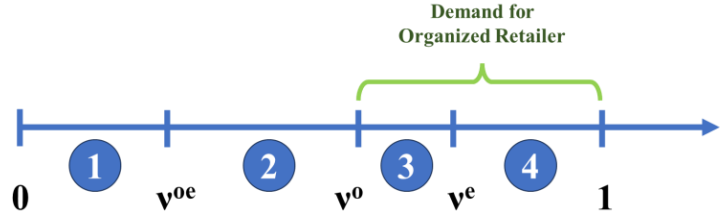

In this case, customers belonging to Zone 1 will not buy as their valuation has not reached the sufficient level for purchase. But, for any customer whose valuation has crossed the valuation indifference point $\nu^{oe}$, the organized retailer is weakly preferred over the e-pharmaceutical retailer. The customers in zone 2, utilities from organized retailer are less than zero. Zone 3 and 4 customers only buy from the organized retailer because utilities from organized retailer is more than e-pharmaceutical retailer. Here, demand for the e-pharmaceutical retailer is zero.

Thus, demand in the unorganized retailer is:

$$D_o = 1 - \nu^o \tag{13}$$

Demand for e-pharmaceutical retailer is:

$$D_e = 0 \tag{14}$$

There is no competition between channels because there is no demand in the e-pharmaceutical retail channel (as shown in Figure 4). Therefore, this setting is not examined.

Similarly, the market setting of the unorganized vs organized retailer and the unorganized retailer vs e-pharmacy is carried out, and the results are shown in Table 2. For ease of readability, the rest of the derivation is presented in Appendix A.

Table 2. Channel Wise Demand for Different Market Scenarios

| Market Setting | Channel wise Demand | | Condition |
|---|---|---|---|
| Organized retailer and E-pharmacy | Demand for organized retailer is: $D_o = 1 - \nu^{oe}$ | Demand for e-pharmaceutical retailer is: $D_e = \nu^{oe} - \nu^e$ | $\nu^e < \nu^o$ |
| | Demand for organized retailer is: $D_o = 1 - \nu^o$ | Demand for e-pharmaceutical retailer is: $D_e = 0$ | $\nu^e > \nu^o$ |
| Unorganized and Organized retailer | Demand for unorganized retailer is: $D_u = 1 - \nu^{uo}$ | Demand for organized retailer is: $D_o = \nu^{uo} - \nu^o$ | $\nu^o < \nu^u$ |
| | Demand for unorganized retailer is: $D_u = 1 - \nu^u$ | Demand for organized retailer is: $D_o = 0$ | $\nu^o > \nu^u$ |
| Unorganized retailer and E-pharmacy | Demand for unorganized retailer is: $D_u = 1 - \nu^{ue}$ | Demand for e-pharmaceutical retailer is: $D_e = \nu^{ue} - \nu^e$ | $\nu^e < \nu^u$ |
| | Demand for unorganized retailer is: $D_u = 1 - \nu^u$ | Demand for e-pharmaceutical retailer is: $D_e = 0$ | $\nu^e > \nu^u$ |

From Table 2, it is shown that only when $\nu^e < \nu^o$, $\nu^o < \nu^u$ and $\nu^e < \nu^u$, all the three channels have demand.

**Proposition 1:** *In a market where unorganized, organized and e-pharmaceutical retailers coexist, the necessary condition for at least two retail channels having demand is as follows:*

$$\nu^e < \nu^o < \nu^u \tag{15}$$

### 4.1.2. Market Where Unorganized Retailer, Organized Retailer and E-Pharmacy Co-Exist:

When customers have the option to purchase from unorganized, organized and e-pharmaceutical retailer, they will compare $U_u$, $U_o$ and $U_e$ to decide which channel to buy from. If $U_u > U_o > U_e$, then the unorganized retailer is preferred over organized retailer then e-pharmaceutical retailer. If $U_o > U_e > U_u$, then the organized retailer is preferred over e-pharmaceutical retailer then unorganized retailer. If $U_e > U_o > U_u$, then the e-pharmaceutical retailer is preferred over organized retailer then unorganized retailer. If $U_u = U_o$ and $U_u = U_e$, customer is indifferent among purchasing from unorganized verses organized and e-pharmaceutical retailers. The indifference customer valuation point $\nu^{uoe}$ is given by following indifference equations:

$$\alpha v^{uoe} - mp_1 - tx = v^{uoe} - mp_2 - \mu_1 \tag{16}$$

$$\alpha v^{uoe} - mp_1 - tx = \theta v^{uoe} - mp_3 - \mu_2 \tag{17}$$

Which simplifies to;

$$v^{uoe} = \frac{m(p_1 - p_2) + (tx - \mu_1) + m(p_1 - p_3) + (tx - \mu_2)}{(\alpha - 1) + (\alpha - \theta)} \tag{18}$$

Which further simplifies to,

$$v^{uoe} = \frac{m(2p_1 - p_2 - p_3) + 2tx - \mu_1 - \mu_2}{2\alpha - \theta - 1} \tag{19}$$

The indifference point $v^{uoe}$ is defined such that if a customer valuation is more than $v^{uoe}$, the customer will prefer unorganized retailer over both organized retailer and e-pharmaceutical retailer. If the customer valuation is less than $v^{uoe}$, then the customer will choose between organized or e-pharmaceutical retailer by comparing $v^{oe}$. Customer valuation when $v^e < v^o$, then only both organized and e-pharmaceutical retailer will have demand (kindly refer to Table 2). Under the conditions of $v^e < v^o < v^u$, there are two possibilities i.e., either $v^{ue} < v^{uo}$ or $v^{ue} > v^{uo}$. Now it is to be found that under which conditions, $v^e < v^o < v^u < v^{uoe}$ and $v^{oe} < v^{uoe}$ holds to have demand in all the three channels (kindly refer to Figure 5).

**Case 1:** when $v^{ue} < v^{uo}$;

$$\frac{m(p_1 - p_3) + tx - \mu_2}{(\alpha - \theta)} < \frac{m(p_1 - p_2) + tx - \mu_1}{(\alpha - 1)} \tag{20}$$

$$[m(p_1 - p_3) + tx - \mu_2]\frac{(\alpha - 1)}{(\alpha - \theta)} < [m(p_1 - p_2) + tx - \mu_1] \tag{21}$$

$$[(mp_1 + tx) - (mp_3 + \mu_2)]\left[\frac{(\alpha - 1) + (\alpha - \theta)}{(\alpha - \theta)} - 1\right] < [(mp_1 + tx) - (mp_2 + \mu_1)] \tag{22}$$

$$[(mp_1 + tx) - (mp_3 + \mu_2)]\left[\frac{(\alpha - 1) + (\alpha - \theta)}{(\alpha - \theta)}\right] < [(mp_1 + tx) - (mp_2 + \mu_1) + (mp_1 + tx) - (mp_3 + \mu_2)] \tag{23}$$

$$\frac{[(mp_1+tx)-(mp_3+\mu_2)]}{(\alpha-\theta)} < \frac{m(p_1-p_2)+(tx-\mu_1)+m(p_1-p_3)+(tx-\mu_2)}{(\alpha-1)+(\alpha-\theta)} \qquad (24)$$

$$v^{ue} < v^{uoe} \qquad (25)$$

As it is shown that, $v^e < v^o < v^u < v^{ue}$ (refer to Table 2), hence

$$v^e < v^o < v^u < v^{ue} < v^{uoe} \qquad (26)$$

Similarly, it can be proved that for case 1,

$$v^{ue} < v^{uoe} < v^{uo} \qquad (27)$$

Now we have to check if $v^{oe} < v^{uoe}$ for this case 1, i.e., $v^{ue} < v^{uoe} < v^{uo}$. We start with equation 25;

$$v^{ue} < v^{uoe}$$

$$\frac{2[(mp_1+tx)-(mp_3+\mu_2)]}{2(\alpha-\theta)} < \frac{[(mp_1+tx)-(mp_2+\mu_1)+(mp_1+tx)-(mp_3+\mu_2)]}{(\alpha-1)+(\alpha-\theta)} \qquad (28)$$

$$2(mp_1+tx)-2(mp_3+\mu_2) < \left[\frac{2(\alpha-\theta)}{(\alpha-1)+(\alpha-\theta)}\right][(mp_1+tx)-(mp_2+\mu_1)+(mp_1+tx)-(mp_3+\mu_2)] \qquad (29)$$

$$[2(mp_1+tx)+(mp_2+\mu_1)-2(mp_3+\mu_2)-(mp_2+\mu_1)] < \left[\frac{2(\alpha-\theta)}{(\alpha-1)+(\alpha-\theta)}\right][2(mp_1+tx)-(mp_2+\mu_1)-(mp_3+\mu_2)] \qquad (30)$$

$$[(mp_2+\mu_1)-(mp_3+\mu_2)] < \left[\frac{2(\alpha-\theta)}{(\alpha-1)+(\alpha-\theta)}-1\right][2(mp_1+tx)-(mp_2+\mu_1)-(mp_3+\mu_2)] \qquad (31)$$

$$[(mp_2+\mu_1)-(mp_3+\mu_2)] < \left[\frac{(1-\theta)}{(\alpha-1)+(\alpha-\theta)}\right][2(mp_1+tx)-(mp_2+\mu_1)-(mp_3+\mu_2)] \qquad (32)$$

$$\frac{[(mp_2+\mu_1)-(mp_3+\mu_2)]}{(1-\theta)} < \frac{[2(mp_1+tx)-(mp_2+\mu_1)-(mp_3+\mu_2)]}{(\alpha-1)+(\alpha-\theta)} \qquad (33)$$

$$v^{oe} < v^{uoe} \qquad (34)$$

Similarly with equation 27, same result will hold. We would like to note that the relative position of $v^u$ and $v^{oe}$ will not change the final demand of the consecutive retail channels. The situation is illustrated in Figure 5.

Figure 5. Customer Valuation When $v^{ue} < v^{uo}$

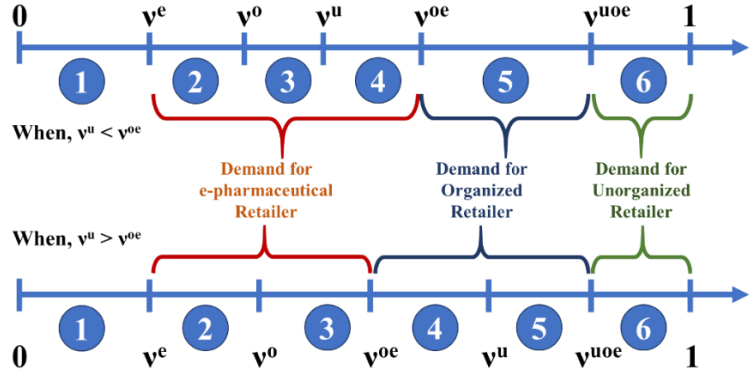

The continuous scale of customer valuation is divided into six zones. For customers in zone 1, utilities from unorganized, organized and e-pharmaceutical retailer are less than zero. Therefore, they will not buy from any of those channels. In zone 2, utility from the e-pharmaceutical retailer $\geq 0$ and it is more than the utility from organized and unorganized retailer in zone 3, so customers will buy from e-pharmaceutical retailer in these two sections. When $v^u < v^{oe}$, customers in zone 4 also purchase from e-pharmaceutical retailer for the same stated reason. But when $v^u > v^{oe}$, the $U_o$ of the customers in zone 4 and 5 is greater than $U_e$ and $U_u$. Hence, they will purchase from organized retailers. Similarly, the customers in zone 5 when $v^u < v^{oe}$ purchase from organized retailers.

Customers will buy from the unorganized retailer only when their valuation exceeds the customer indifference point $v^{uoe}$ in zone 6 where $U_u > U_o$ and $U_u > U_e$.

Demand for unorganized retail channel is:

$$D_u = 1 - v^{uoe} = 1 - \frac{m(2p_1 - p_2 - p_3) + 2tx - \mu_1 - \mu_2}{2\alpha - \theta - 1} \tag{35}$$

Demand for organized retail channel is:

$$D_o = v^{uoe} - v^{oe} = \frac{m(2p_1 - p_2 - p_3) + 2tx - \mu_1 - \mu_2}{2\alpha - \theta - 1} - \frac{m(p_2 - p_3) + \mu_1 - \mu_2}{(1 - \theta)} \tag{36}$$

Demand for e-pharmaceutical retail channel is:

$$D_e = v^{oe} - v^e = \frac{m(p_2 - p_3) + \mu_1 - \mu_2}{(1 - \theta)} - \frac{mp_3 + \mu_2}{\theta} \qquad (37)$$

**Case 2:** when $v^{ue} > v^{uo}$; similar way it can be proved that (refer to Appendix A),

$$v^{uo} < v^{uoe} < v^{ue} \qquad (38)$$

But,

$$v^{oe} > v^{uoe} \qquad (39)$$

The situation is illustrated in Figure 6.

Figure 6. Customer Valuation When $v^{ue} > v^{uo}$

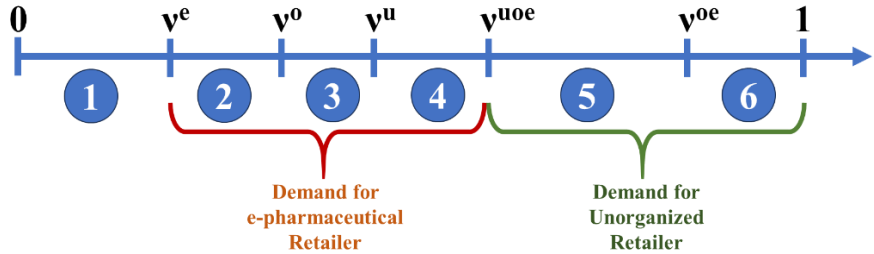

There is no competition among the three channels because there is no demand in the organized retail channel (as shown in Figure 6). Therefore, this setting is not examined.

## 4.2. Derivation of the Profit Functions

In the context of Bertrand price competition, unorganized retailers, organized retailers, and e-pharmaceutical retailers fix their pricing with the goal of maximizing their individual profits. Therefore, we proceed by considering the profit function of the three retail channels. The profit function of each retail channel can be estimated by multiplying the margin for each product sold by its respective demand. The margin for each product sold is calculated by deducting the marginal cost from the price of the product. From equations (35), (36), and (37), the profit function of the unorganized retail channel is:

$$\pi_1 = \beta(p_1 - c_1)D_u = \beta(p_1 - c_1)\left[1 - \frac{m(2p_1 - p_2 - p_3) + 2tx - \mu_1 - \mu_2}{2\alpha - \theta - 1}\right] \qquad (40)$$

Similarly, the profit function of organized retail channel is:

$$\pi_2 = \beta(p_2 - c_2)D_o = \beta(p_2 - c_2)\left[\frac{m(2p_1 - p_2 - p_3) + 2tx - \mu_1 - \mu_2}{2\alpha - \theta - 1} - \frac{m(p_2 - p_3) + \mu_1 - \mu_2}{(1-\theta)}\right] \quad (41)$$

The profit function of e-pharmaceutical retail channel is:

$$\pi_3 = \beta(p_3 - c_3)D_e = \beta(p_3 - c_3)\left[\frac{m(p_2 - p_3) + \mu_1 - \mu_2}{(1-\theta)} - \frac{mp_3 + \mu_2}{\theta}\right] \quad (42)$$

## 4.3. Derivation of the Optimal Prices

In Bertrand price competition, organized, unorganized and e-pharmaceutical retailers declare their respective channel prices. Since the prices are declared simultaneously, none of the retail channels is in a dominant position.

Optimal prices are calculated by finding the first-order conditions of $\pi_i, i \in \{1,2,3\}$ with respect to $p_i, i \in \{1,2,3\}$ and then by simultaneously solving them. $\frac{\partial \pi_1}{\partial p_1} = 0$, $\frac{\partial \pi_2}{\partial p_2} = 0$ and $\frac{\partial \pi_3}{\partial p_3} = 0$.

**Lemma 1.** The optimal prices of the three pharmaceutical retail channels are given below. The condition for optimality is as follows:

$$\alpha > \frac{1+\theta}{2} \quad (43)$$

Optimal price for unorganized pharmaceutical retail channel:

$$p_1^* = \frac{1}{m(2 + 4\alpha(-4 + \theta) + 11\theta - \theta^2)}\begin{pmatrix} -2tx + 4\alpha + 8tx\alpha - 8\alpha^2 - 3\theta - 5tx\theta + 9\alpha\theta - 2tx\alpha\theta + 2\alpha^2\theta - 3\theta^2 + \\ tx\theta^2 - \alpha\theta^2 + 2m(\alpha(-4 + \theta) + 3\theta)c_1 - m(\alpha - \theta)(2 + \theta)c_2 + mc_3 - 3m\alpha c_3 + \\ 2m\theta c_3 - 2\alpha\mu_1 + 2\theta\mu_1 - \alpha\theta\mu_1 + \theta^2\mu_1 + \mu_2 - 3\alpha\mu_2 + 2\theta\mu_2 \end{pmatrix}$$

$$(44)$$

Optimal price for organized pharmaceutical retail channel:

$$p_2^* = \frac{1}{m\left(2 + 4\alpha\left(-4+\theta\right)+11\theta-\theta^2\right)}\begin{pmatrix} 2 - 4tx - 4\alpha + 4tx\theta + 4\alpha\theta - 2\theta^2 + 4m\left(-1+\theta\right)c_1 + \\ 8m\left(-\alpha+\theta\right)c_2 + 3mc_3 - 4m\alpha c_3 + m\theta c_3 - 2\mu_1 + \\ 8\alpha\mu_1 - 3\theta\mu_1 - 4\alpha\theta\mu_1 + \theta^2\mu_1 + 3\mu_2 - 4\alpha\mu_2 + \theta\mu_2 \end{pmatrix} \quad (45)$$

Optimal price for e-pharmaceutical retail channel:

$$p_3^* = \frac{1}{m\left(2 + 4\alpha\left(-4+\theta\right)+11\theta-\theta^2\right)}\begin{pmatrix} \theta - 2tx\theta - 2\alpha\theta + 2tx\theta^2 + 2\alpha\theta^2 - \theta^3 + 2m\left(-1+\theta\right)\theta c_1 + \\ 4m\theta\left(-\alpha+\theta\right)c_2 + mc_3 - 8m\alpha c_3 + 7m\theta c_3 - 4\alpha\theta\mu_1 + \\ 4\theta^2\mu_1 - \mu_2 + 8\alpha\mu_2 - 4\theta\mu_2 - 4\alpha\theta\mu_2 + \theta^2\mu_2 \end{pmatrix} \quad (46)$$

Lemma 1 proves the existence and uniqueness of the equilibrium. For ease of readability, the proof of Lemma 1 is given in Appendix B.

## 5. Numerical Simulation

In this section, a numerical simulation is carried out using real data from a field study.

### 5.1. Base Case Scenario

Our field study focused on a block of the eastern state of India, West Bengal. A semi-structured interview of unorganized pharmaceutical retailers, consumers, and stakeholders in the study area is carried out. The following parameter values for numerical simulation are collected from literature and the semi-structured interview: $\alpha = 0.9$, $\theta = 0.6$, $\beta = 1$, $m = 1$, $t = 10$, $x = 0.86$, $c_1 = 175$, $c_2 = 140$, $c_3 = 140$, $\mu_1 = 20$, $\mu_2 = 20$. The results are shown in Table 3. Distance to the nearest unorganized pharmaceutical retailer for a customer (x) is estimated from the map by calculating the average distance of each customer from their individual nearest unorganized pharmaceutical retailer store.

Table 3. Base Case Scenario Results

|  | Unorganized pharmaceutical retailers | Organized pharmaceutical retailers | E-pharmaceutical retailers |
|---|---|---|---|
| **Optimal price** | 158.26 | 149.56 | 110.87 |
| **Optimal profit** | 2803.11 | 685.40 | 3536.14 |

## 5.2. Sensitivity Analysis

The first step in sensitivity analysis is to find the range of each parameter and select a proper distribution function. Table 4 shows the parameters with their range and distribution. After that, the one-factor-at-a-time (OFAT) approach is carried out for each parameter, and the model is simulated by changing the input parameter values.

Table 4. Parameter Distribution for Sensitivity Analysis

| Sl no. | Parameters | Range | Distribution |
|--------|-----------|-------|--------------|
| 1 | Customer acceptance of unorganized pharmaceutical retailer ($\alpha$) | (0.8, 1] | Uniform |
| 2 | Customer acceptance of online e-pharmaceutical retailer ($\theta$) | (0, 0.8) | Uniform |
| 3 | Transportation cost per unit distance incurred by the customer ($t$) | [0, 20] | Uniform |
| 4 | Distance to the nearest unorganized pharmaceutical retailer for a customer ($x$) | [0, 10] | Uniform |
| 5 | Probability of product category level demand ($\beta$) | (0, 1] | Uniform |
| 6 | Marginal utility of money ($m$) | (0, 1] | Uniform |
| 7 | Customer disutility of purchasing from organized pharmaceutical retailers ($\mu_1$) | [0, 200] | Uniform |
| 8 | Customer disutility of purchasing from e-pharmaceutical retailers ($\mu_2$) | [0, 200] | Uniform |
| 9 | Marginal cost incurred by the unorganized pharmaceutical retailers ($c_1$) | [119, 231] | Uniform |
| 10 | Marginal cost incurred by the organized pharmaceutical retailers ($c_2$) | [49, 231] | Uniform |
| 11 | Marginal cost incurred by the e-pharmaceutical retailers ($c_3$) | [49, 231] | Uniform |

## 6. Results and Discussions

The sensitivity analysis provides valuable insights on the optimal pricing strategies (Figure 7) across all three retail channels. The results of the sensitivity analysis for optimal profit are depicted in Figure C in Appendix C.

Figure 7. Results Of Sensitivity Analysis for Optimal Price

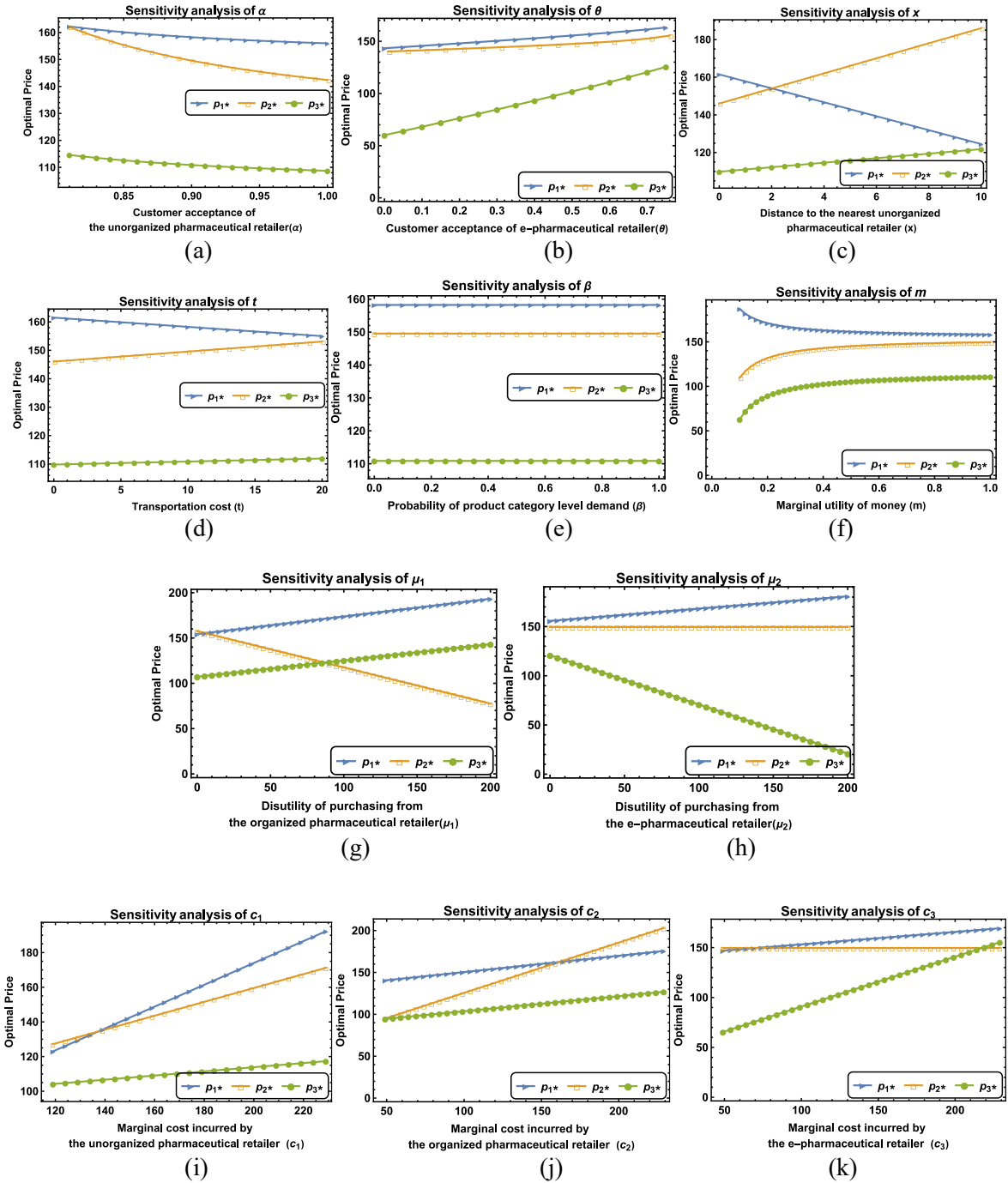

The analysis highlights how various factors affect optimal pricing decisions, leading to the following findings:

### 6.1. Effect of Customer Acceptance of Unorganized Retailers (*α*)

As customer acceptance of unorganized pharmaceutical retailers (*α*) increases, the preference for unorganized retailers grows. A counterintuitive behavior is evidenced here that with the increase in '*α*' along with an increase in discounts (Figure 7(a)), will result in a larger fall in optimal profits of both unorganized and organized pharmaceutical retailers (Figure C(a)). This counterintuitive behavior requires further exploration. Interestingly, an increase in '*α*' is found to benefit the e-pharmaceutical retailers more in terms of their profits (Figure C(a)).

### 6.2. Effect of Customer Acceptance of Online E-Pharmaceutical Retailers (*θ*)

The sensitivity analysis shows that as customer acceptance of e-pharmaceutical retailers (*θ*) increases, they have reduced the discount margins, thereby increasing the prices (Figure 7(b)). This increase in price initially provided the e-retailer with a profit advantage, which finally resulted in the collapse of their profit margin beyond the '*θ*' value of 0.2 (Figure C(b)). Moreover, even though the market share of e-retailers has fallen beyond the value 0.2 (*θ*), the optimal profit of unorganized and organized pharmaceutical retailers is found to increase only beyond the '*θ*' value of 0.6. Hence, it is evidenced that beyond the value 0.6 (*θ*), there is an 'overshoot' in the optimal profit and thus the market share of organized and unorganized pharmaceutical retailer channels (Figure C(b)).

### 6.3. Effect of Distance to Unorganized Retailers (*x*) and Effect of Transportation Cost (*t*)

As the distance to the nearest unorganized pharmaceutical retailer (*x*) increases, customer utility from these unorganized retailers decreases due to the inconvenience of traveling farther. In response, unorganized retailers have reduced their prices (i.e., more discounts) to compensate for the decreased utility, thereby attracting customers towards increasing their competitive advantage (Figure 7(c)). Interestingly, a counterintuitive behavior is evidenced here as low prices finally resulted in exponentially increasing profits for the unorganized pharmaceutical retailers (Figure C(c)). This behavior can be attributed to the customer footfalls outpacing the price discounts offered by unorganized retailers. A similar behavior is also evidenced in the case of an increase in transportation cost per unit distance (*t*) (Figures 7(d) and C(d)).

### 6.4. Effect of Product Category Level Demand (*β*)

Jerath et al. (2016) from their study observed that the probability of product category level demand has a negative influence on the price charged by the retailers. Unlike Jerath et al.'s study, the present analysis found that the probability of product category level demand (*β*) does not influence the optimal pricing strategies. From the analysis, it is evident that as the '$\beta$' value increases, the optimal profits for all three retail channels will also increase even when the price remains stabilized (Figures 7(e) and C(e)).

## 6.5. Effect of Marginal Utility of Money ($m$)

A counterintuitive relationship is evidenced between the marginal utility of money and the optimal price. Unlike economic theory, the analysis exhibits that the marginal utility of money has no influence on the optimal price (Figure 7(f)), but does have an influence on the optimal profit of the retailers. Hence it is inferred that as the marginal utility of money is increasing for consumers coupled with low sensitivity to price changes, they are more likely to make purchases which in turn leads to the increase in retailers' sales revenue and profit as evidenced from Figure C(f).

## 6.6. Effect of Customer Disutility of Purchasing from the Organized Retailers ($\mu_1$) and from E-Pharmaceutical Retailers ($\mu_2$)

As disutility of purchasing from organized pharmaceutical retailers ($\mu_1$) increases, the organized retailers become less preferred by the customers. In response the organized retailers will sacrifice their discount margins thereby reducing the prices to attract more customers. This will in turn result in attracting more customers leading to increased sales revenue and profits for the organized retailers (Figures 7(g) and C(g)). This behavior of the organized retailers will finally affect the profit margins of the unorganized and e-retailers (Figure C(g)). A similar behavioral dynamic is evidenced in the case of the disutility of purchasing from e-pharmaceutical retailers ($\mu_2$) (Figures 7(h) and C(h)).

## 6.7. Effect of Marginal Costs ($c_1, c_2, c_3$)

As the marginal costs ($c_1$ for unorganized, $c_2$ for organized, and $c_3$ for e-pharmaceutical retailers) increase for each retail channel, the corresponding channel should reduce its discount margins in order to raise its prices proportionally to maintain profit levels. This ensures that rising operational costs are accounted for without sacrificing profitability. The same is evidenced in the results of sensitivity analysis for all three retail channels (see Figures 7(i) and C(i); Figures 7(j) and C(j); and Figures 7(k) and C(k) for unorganized, organized and e-pharmaceutical retailers respectively).

## 7. Conclusions

This study aimed to investigate the effect of customer choices and different market conditions on pricing strategies, focusing on organized, unorganized, and e-pharmaceutical retail channels in India. Accordingly, a block in the eastern state of India is selected as the study location. Based on an extensive literature review and in-depth interviews, we have identified key factors that influence how customers choose their preferred retail channels, and corresponding utility functions are then developed analytically. After that, a game theoretical analysis of organized, unorganized, and e-pharmaceutical retailers is conducted, and finally, a sensitivity analysis is carried out.

The sensitivity analysis reveals that customer acceptance of unorganized pharmaceutical retailers ($\alpha$), customer acceptance of e-pharmaceutical retailers ($\theta$), distance to the nearest unorganized pharmaceutical retailer ($x$), transportation cost per unit distance incurred by the customer ($t$), disutility of purchasing from organized chain pharmaceutical retailers ($\mu 1$), disutility of purchasing from online e-pharmaceutical retailers ($\mu 2$) and marginal costs ($c1$, $c2$, $c3$) are key determinants of both the optimal pricing and profit of the triple-channel pharmaceutical retail supply chain. While the product category level demand ($\beta$) and marginal utility of money ($m$) did not influence pricing strategies, they do have an influence on the optimal profit of these retailers. These findings suggest that retailers must adopt dynamic pricing strategies based on customer preferences and market conditions. Specifically, the growing acceptance of e-pharmaceutical retailers justifies price increases, while unorganized retailers should maintain low prices despite increased demand to remain competitive. Our model provides insights into certain empirical observations regarding pharmaceutical retailing in developing economies, which involves the factors that are influential to the long-term survival of unorganized pharmaceutical retailers amidst the rise of organized and e-retailing in India. Our research findings offer valuable insights for policymakers facing challenges in achieving a balanced growth among the organized, unorganized, and e-pharmaceutical retail sectors in emerging economies.

The findings of this research are based on a few assumptions regarding the behavior of the customer and the conditions of the market, which may restrict their applicability in different regions or under different market environments. Future research could further explore similar dynamics in other geographical settings. Future researchers could explore additional variables, such as the impact of digital marketing or changing regulations, and the impact of word-of-mouth to refine these pricing strategies. Additional insights could be obtained by conducting a more thorough examination of consumer preferences among a variety of demographic groups. In India's burgeoning retail market, this research will shed light on the relevance of the unorganized pharmaceutical retail sector. This research will help the organized, unorganized, and e-retail sectors design policies to maximize gains from the booming Indian consumer market and boost the economy.

## Appendix A: Derivation of the Demand Function:

### A.1. Market Where Unorganized and Organized Retailers are Present:

When customers have the option to purchase from unorganized and organized retailer, they will compare $U_u$ and $U_o$. If $U_u > U_o$, then the unorganized retailer is preferred over organized retailer and if $U_u = U_o$, customer is indifferent between purchasing from unorganized and organized retailers. The indifference customer valuation point $v^{uo}$ is given by following indifference equation:

$$\alpha v^{uo} - mp_1 - tx = v^{uo} - mp_2 - \mu_1 \qquad (1)$$

Which simplifies to;

$$v^{uo} = \frac{m(p_1 - p_2) + tx - \mu_1}{(\alpha - 1)} \qquad (2)$$

Note: if the valuation exceeds $v^{uo}$, customer will prefer the unorganized retailer over organized retailer.

Now we have two cases: $v^o < v^u$ and $v^o > v^u$.

**Case 1:** when $v^o < v^u$,

$$mp_2 + \mu_1 < \frac{mp_1 + tx}{\alpha} \qquad (3)$$

$$mp_2 + \mu_1 < (mp_1 + tx)(1 - \frac{\alpha - 1}{\alpha}) \qquad (4)$$

$$\frac{(\alpha - 1)}{\alpha}(mp_1 + tx) < mp_1 + tx - mp_2 - \mu_1 \qquad (5)$$

$$\frac{mp_1 + tx}{\alpha} < \frac{m(p_1 - p_2) + tx - \mu_1}{(\alpha - 1)} \qquad (6)$$

Or, $v^u < v^{uo}$, thus $v^o < v^u < v^{uo}$. Graphically, the situation is shown in Figure A1.

Figure A1. Customer Valuation When $v^o < v^u$

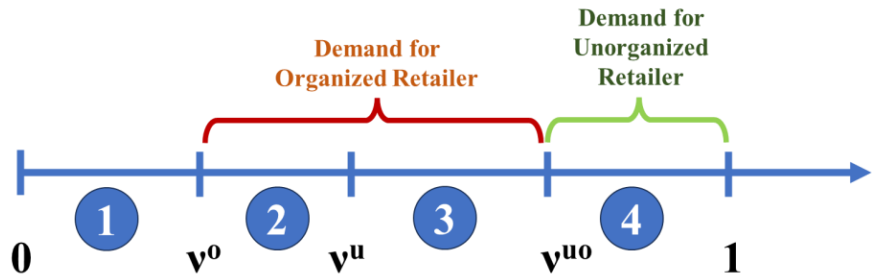

The continuous scale of customer valuation is divided into four zones, namely 1, 2, 3 and 4. For customers in zone 1, both utilities from unorganized and organized retailer are less than zero. Therefore, they will not buy from either channel. In zone 2 utility from the organized retailer $\geq 0$ and in zone 3, the utility from the organized retailer is more than the utility from unorganized retailer, so customers will buy from organized retailer in these two sections. Customers will buy from the unorganized retailer only when their valuation exceeds the customer indifference point $v^{uo}$ i.e., in zone 4 where $U_u > U_o$.

Demand for unorganized retailer is:

$$D_u = 1 - v^{uo} \tag{7}$$

Demand for organized retailer is:

$$D_o = v^{uo} - v^{o} \tag{8}$$

**Case 2:** when $v^o > v^u$,

$$mp_2 + \mu_1 > \frac{mp_1 + tx}{\alpha} \tag{9}$$

$$mp_2 + \mu_1 > (mp_1 + tx)(1 - \frac{\alpha - 1}{\alpha}) \tag{10}$$

$$\frac{(\alpha - 1)}{\alpha}(mp_1 + tx) > mp_1 + tx - mp_2 - \mu_1 \tag{11}$$

$$\frac{mp_1 + tx}{\alpha} > \frac{m(p_1 - p_2) + tx - \mu_1}{(\alpha - 1)} \tag{12}$$

Or, $v^u > v^{uo}$, thus $v^o > v^u > v^{uo}$. Graphically, the situation is shown in Figure A2.

Figure A2. Customer Valuation When $v^o > v^u$

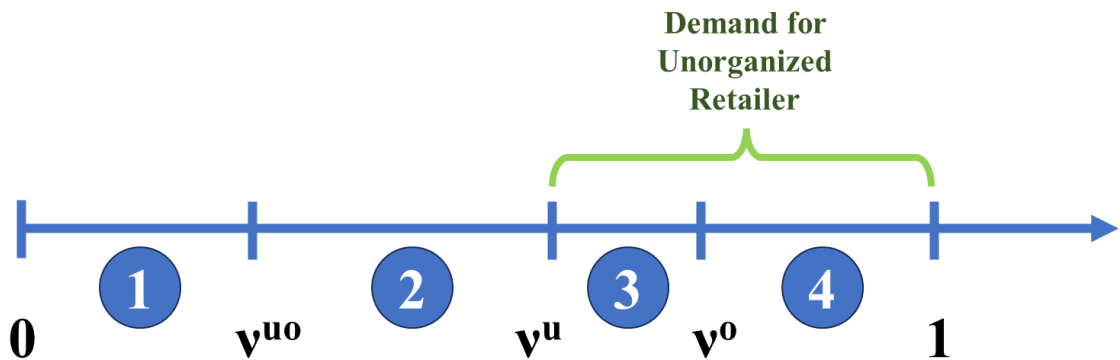

In this case, customers belonging to Zone 1 will not buy as their valuation has not reached the sufficient level for purchase. But, for any customer whose valuation has crossed the valuation indifference point $v^{uo}$, the unorganized retailer is weakly preferred over the organized retailer. The customers in zone 2, utilities from organized and unorganized retailers are less than zero. Zone 3 and 4 customers only buy from the unorganized retailer because utilities from unorganized retailer is more than organized retailer. Here, demand for the organized retailer is zero.

Thus, demand in the unorganized retailer is:

$$D_u = 1 - v^u \tag{13}$$

Demand for organized retailer is:

$$D_o = 0 \tag{14}$$

As there is no demand in the organized retail channel, there is no competition between channels, and this setting is not analysed.

## A.2. Market Where Unorganized Retailer and E-Pharmacy are Present:

When customers have the option to purchase from unorganized and e-pharmaceutical retailer, they will compare $U_u$ and $U_e$. If $U_u > U_e$, then the unorganized retailer is preferred over e-pharmaceutical retailer and if $U_u = U_e$, customer is indifferent between purchasing from unorganized and e-pharmaceutical retailers. The indifference customer valuation point $v^{ue}$ is given by following indifference equation:

$$\alpha v^{ue} - mp_1 - tx = \theta v^{ue} - mp_3 - \mu_2 \tag{15}$$

Which simplifies to;

$$v^{ue} = \frac{m(p_1 - p_3) + tx - \mu_2}{(\alpha - \theta)} \tag{16}$$

Note: if the valuation exceeds $v^{ue}$, customer will prefer the unorganized retailer over e-pharmaceutical retailer.

Now we have two cases: $v^e < v^u$ and $v^e > v^u$.

**Case 1:** when $v^e < v^u$,

$$\frac{mp_3 + \mu_2}{\theta} < \frac{mp_1 + tx}{\alpha} \tag{17}$$

$$mp_3 + \mu_2 < (mp_1 + tx)(1 - \frac{\alpha - \theta}{\alpha}) \tag{18}$$

$$\frac{(\alpha - \theta)}{\alpha}(mp_1 + tx) < mp_1 + tx - mp_3 - \mu_2 \tag{19}$$

$$\frac{mp_1 + tx}{\alpha} < \frac{m(p_1 - p_3) + tx - \mu_2}{(\alpha - \theta)} \tag{20}$$

Or, $v^u < v^{ue}$, thus $v^e < v^u < v^{ue}$. Graphically, the situation is shown in Figure A3.

Figure A3. Customer Valuation When $v^e < v^u$

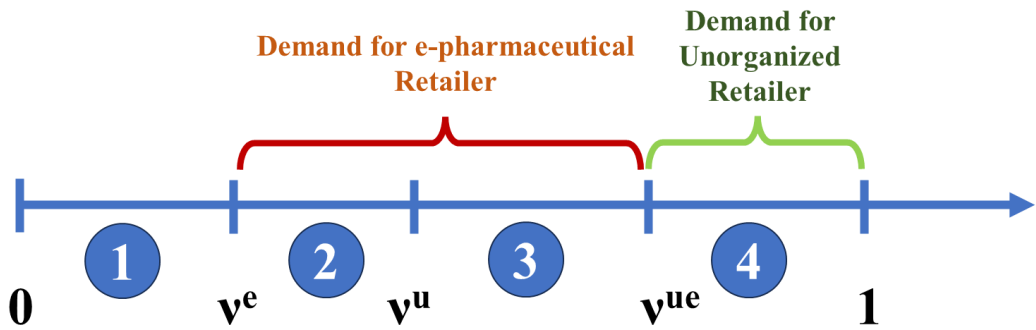

The continuous scale of customer valuation is divided into four zones, namely 1, 2, 3 and 4. For customers in zone 1, both utilities from unorganized and e-pharmaceutical retailer are less than zero. Therefore, they will not buy from either channel. In zone 2, utility from the e-pharmaceutical retailer ≥ 0 and it is more than the utility from unorganized retailer in zone 3, so customers will buy from e-pharmaceutical retailer in these two sections. Customers will buy from the unorganized retailer only when their valuation exceeds the customer indifference point $v^{ue}$ in zone 4 where $U_u > U_e$.

Demand for unorganized retailer is:

$$D_u = 1 - v^{ue} \tag{21}$$

Demand for e-pharmaceutical retailer is:

$$D_e = v^{ue} - v^e \tag{22}$$

**Case 2:** when $v^e > v^u$,

$$\frac{mp_3 + \mu_2}{\theta} > \frac{mp_1 + tx}{\alpha} \tag{23}$$

$$mp_3 + \mu_2 > (mp_1 + tx)(1 - \frac{\alpha - \theta}{\alpha}) \qquad (24)$$

$$\frac{(\alpha - \theta)}{\alpha}(mp_1 + tx) > mp_1 + tx - mp_3 - \mu_2 \qquad (25)$$

$$\frac{mp_1 + tx}{\alpha} > \frac{m(p_1 - p_3) + tx - \mu_2}{(\alpha - \theta)} \qquad (26)$$

Or, $v^u > v^{ue}$, thus $v^e > v^u > v^{ue}$. Graphically, the situation is shown in Figure A4.

Figure A4. Customer Valuation When $v^e > v^u$

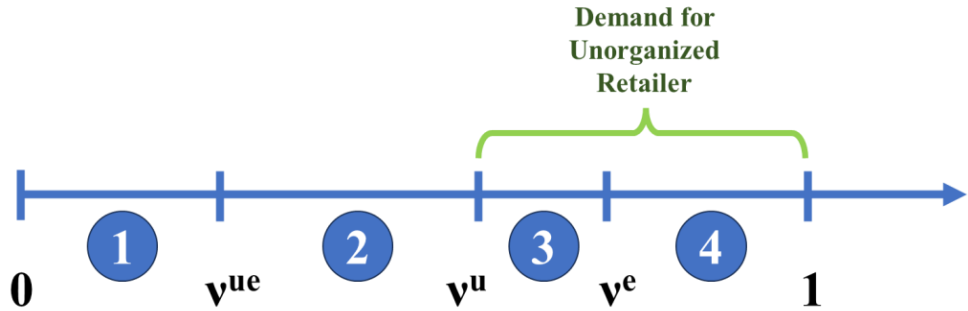

In this case, customers belonging to Zone 1 will not buy as their valuation has not reached the sufficient level for purchase. But, for any customer whose valuation has crossed the valuation indifference point $v^{ue}$, the unorganized retailer is weakly preferred over the e-pharmaceutical retailer. The customers in zone 2, utilities from unorganized retailer are less than zero. Zone 3 and 4 customers only buy from the unorganized retailer because utilities from unorganized retailer is more than e-pharmaceutical retailer. Here, demand for the e-pharmaceutical retailer is zero.

Thus, demand in the unorganized retailer is:

$$D_u = 1 - v^u \qquad (27)$$

Demand for e-pharmaceutical retailer is:

$$D_e = 0 \qquad (28)$$

As there is no demand in the e-pharmaceutical retail channel, there is no competition between channels, and this setting is not analysed.

### A.3. Market Where Unorganized Retailer, Organized Retailer and E-Pharmacy Co-Exist:

**Case 2:** when $v^{ue} > v^{uo}$ ; similar way it can be proved that,

$$v^{uo} < v^{uoe} < v^{ue} \tag{29}$$

But,

$$v^{oe} > v^{uoe} \tag{30}$$

Graphically, the situations are shown in Figure A5.

Figure A5. Customer Valuation When $v^{ue} > v^{uo}$

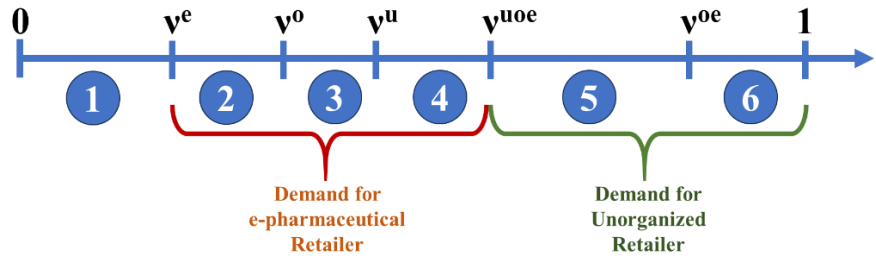

The continuous scale of customer valuation is divided into six zones. For customers in zone 1, utilities from unorganized, organized and e-pharmaceutical retailer are less than zero. Therefore, they will not buy from any of those channels. In zone 2, utility from the e-pharmaceutical retailer $\geq 0$ and it is more than the utility from organized and unorganized retailer in zone 3 and 4 respectively, so customers will buy from e-pharmaceutical retailer in these two sections. Customers will buy from the unorganized retailer only when their valuation exceeds the customer indifference point $v^{uoe}$ in zone 5 and 6 where $U_u > U_o$ and $U_u > U_e$ .

Demand for unorganized retailer is:

$$D_u = 1 - v^{uoe} = 1 - \frac{m(2p_1 - p_2 - p_3) + 2tx - \mu_1 - \mu_2}{2\alpha - \theta - 1} \tag{31}$$

Demand for organized retailer is:

$$D_o = 0 \tag{32}$$

Demand for e-pharmaceutical retailer is:

$$D_e = v^{uoe} - v^e = \frac{m(2p_1 - p_2 - p_3) + 2tx - \mu_1 - \mu_2}{2\alpha - \theta - 1} - \frac{mp_3 + \mu_2}{\theta} \tag{33}$$

As there is no demand in the organized retail channel, there is no competition between channels, and hence this setting is not analysed.

**Appendix B: Determination of Nash Equilibrium:**

The profit function of the unorganized retail channel is:

$$\pi_1 = \beta(p_1 - c_1)D_u = \beta(p_1 - c_1)\left[1 - \frac{m(2p_1 - p_2 - p_3) + 2tx - \mu_1 - \mu_2}{2\alpha - \theta - 1}\right] \tag{34}$$

The profit function of the organized retail channel is:

$$\pi_2 = \beta(p_2 - c_2)D_o = \beta(p_2 - c_2)\left[\frac{m(2p_1 - p_2 - p_3) + 2tx - \mu_1 - \mu_2}{2\alpha - \theta - 1} - \frac{m(p_2 - p_3) + \mu_1 - \mu_2}{(1 - \theta)}\right] \tag{35}$$

The profit function of the e-pharmaceutical retail channel is:

$$\pi_3 = \beta(p_3 - c_3)D_e = \beta(p_3 - c_3)\left[\frac{m(p_2 - p_3) + \mu_1 - \mu_2}{(1 - \theta)} - \frac{mp_3 + \mu_2}{\theta}\right] \tag{36}$$

First, we calculated the first-order conditions of (34), (35), and (36) with respect to $p_i, i \in \{1, 2, 3\}$, which are given in (37) through (39) respectively. These are the necessary conditions.

$$\frac{4mp_1}{2\alpha - \theta - 1} - \frac{mp_2}{2\alpha - \theta - 1} - \frac{mp_3}{2\alpha - \theta - 1} + \frac{2tx - \mu_1 - \mu_2 - 2mc_1}{2\alpha - \theta - 1} - 1 = 0 \tag{37}$$

$$\frac{2mp_1}{2\alpha - \theta - 1} - \left[\frac{1}{2\alpha - \theta - 1} + \frac{1}{1 - \theta}\right]2mp_2 - \left[\frac{1}{2\alpha - \theta - 1} - \frac{1}{1 - \theta}\right]mp_3 + \frac{2tx - \mu_1 - \mu_2 + mc_2}{2\alpha - \theta - 1} + \frac{mc_2 - \mu_1 + \mu_2}{1 - \theta} = 0 \tag{38}$$

$$\frac{m}{1 - \theta}p_2 - \left[\frac{1}{1 - \theta} + \frac{1}{\theta}\right]2mp_3 + \frac{\mu_1 - \mu_2 + mc_3}{1 - \theta} + \frac{mc_3 - \mu_2}{\theta} = 0 \tag{39}$$

Next, we find the second-order derivatives of (34), (35), and (36) with respect to $p_i, i \in \{1, 2, 3\}$, which are given in (40) through (42) respectively. These are the sufficient conditions.

$$\frac{\partial^2 \pi}{\partial p_1^2} = -\frac{4m\beta}{2\alpha - \theta - 1} \tag{40}$$

$$\frac{\partial^2 \pi_2}{\partial p_2^2} = -2m\beta\left[\frac{1}{2\alpha - \theta - 1} - \frac{1}{1 - \theta}\right] \tag{41}$$

$$\frac{\partial^2 \pi_3}{\partial p_3^2} = -2m\beta\left[\frac{1}{1 - \theta} + \frac{1}{\theta}\right] \tag{42}$$

From (40), (41) and (42), the conditions for the profit functions to be concave, i.e., $\frac{\partial^2 \pi_1}{\partial p_1^2} < 0$, $\frac{\partial^2 \pi_2}{\partial p_2^2} < 0$ and $\frac{\partial^2 \pi_3}{\partial p_3^2} < 0$ are as follows:

$$\alpha > \frac{1 + \theta}{2} \text{ and } \alpha > \theta \tag{43}$$

## Appendix C: Sensitivity Analysis Results for Optimal Profit

Figure C. Results of Sensitivity Analysis for Optimal Profit

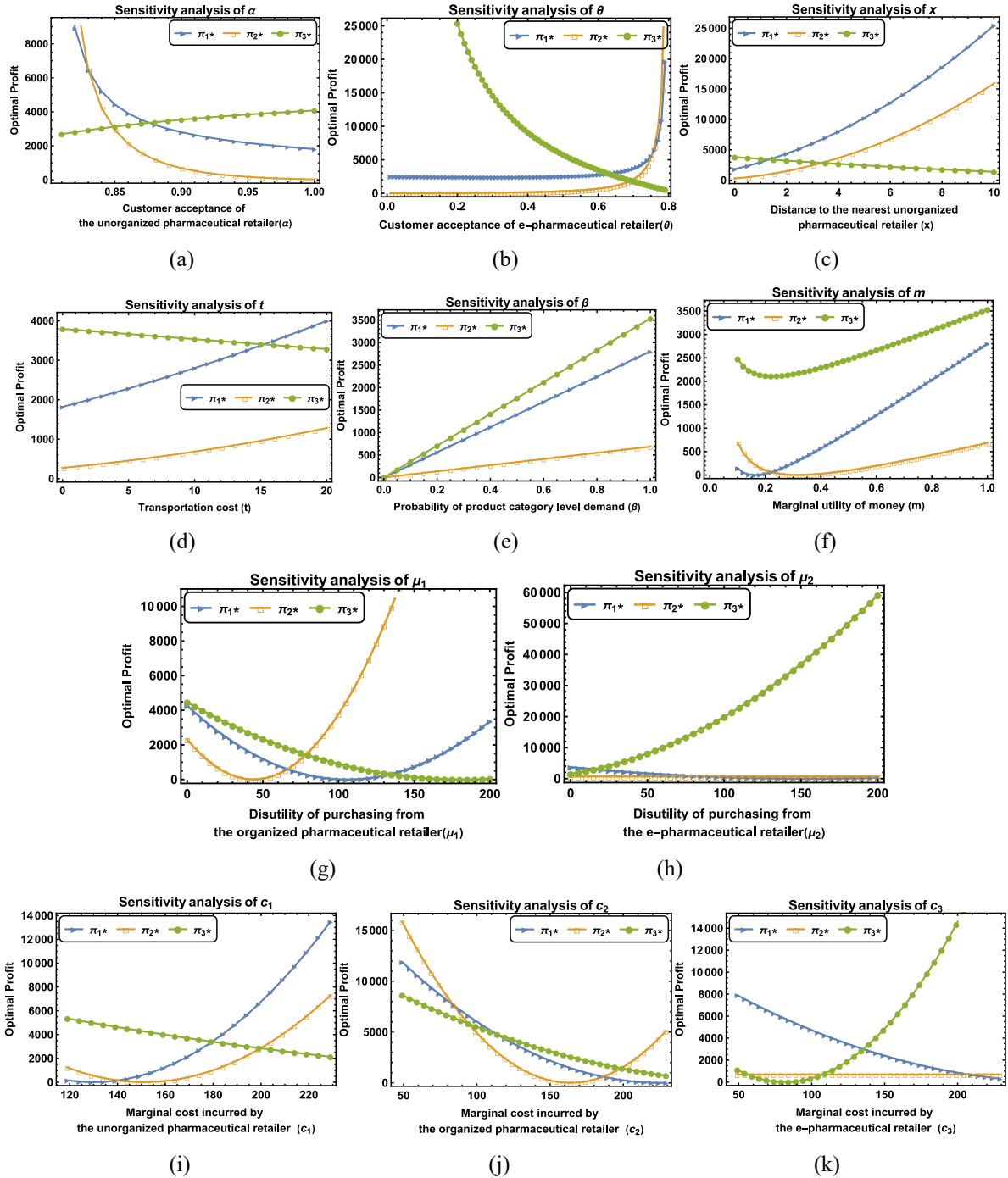

(a)   (b)   (c)

(d)   (e)   (f)

(g)   (h)

(i)   (j)   (k)

Figure C represents the results of the sensitivity analysis of the optimal profit of unorganized, organized, and e-pharmaceutical retail channels.